\documentclass[12pt,preprint]{aastex}
\begin{document}

\title{Binarity as a key factor in protoplanetary disk evolution: {\it Spitzer\,} disk census of the $\eta$ Chamaeleontis cluster}

\author{
J. Bouwman$^{1}$, W. A. Lawson$^{2}$, C. Dominik$^{3}$, 
E. D. Feigelson$^{5}$, Th. Henning$^{1}$, 
A. G. G. M. Tielens$^{6}$, L. B. F. M. Waters$^{3,4}$
}

\altaffiltext{1}{Max Planck Institute for Astronomy, K\"onigstuhl 
17, D-69117 Heidelberg, Germany}

\altaffiltext{2}{School of Physical, Environmental and Mathematical 
Sciences, University of New South Wales, Australian Defence Force 
Academy, Canberra ACT 2600, Australia}

\altaffiltext{3}{Astronomical Institute "Anton Pannekoek", 
University of Amsterdam, Kruislaan 403 NL-1098 SJ Amsterdam, 
Netherlands}

\altaffiltext{4}{Instituut voor Sterrenkunde, KU Leuven, Celestijnenlaan 200B, 3001 Leuven, Belgium}

\altaffiltext{5}{Department of Astronomy \& Astrophysics, 
Pennsylvania State University, University Park PA 16802}

\altaffiltext{6}{Kapteyn Astronomical Institute, University 
of Groningen, PO Box 800, 9700 AV Groningen, Netherlands}


\begin{abstract}
The formation of planets is directly linked to the evolution of
the circumstellar (CS) disk from which they are born. The dissipation timescales of
CS disks are, therefore, of
direct astrophysical importance in evaluating the time available for planet
formation. We employ {\it Spitzer Space Telescope\,} spectra to complete the 
CS disk census for the late-type members
of the $\simeq 8$ Myr-old $\eta$ Chamaeleontis star cluster. Of the 15 K- and
M-type members, eight show excess emission. We find that the presence of
a CS disk is anti-correlated with binarity, with all but one disk associated with
single stars. With nine single stars in total, about 80\% retain a CS disk.
Of the six known or suspected close binaries the only CS disk is associated with the
primary of RECX 9. No circumbinary disks have been detected. We also find 
that stars with disks are slow rotators with surface values
of specific angular momentum $j = 2-15 j_{\odot}$.  All high specific angular
momentum systems with $j = 20-30 j_{\odot}$ are confined to the primary stars of
binaries. This provides novel empirical evidence for rotational
disk locking and again demonstrates the much shorter disk lifetimes in close binary
systems compared to single star systems. We estimate the
characteristic mean disk dissipation timescale to be $\sim 5$
Myr and $\approx 9$ Myr for the binary and single star systems, respectively.

\end{abstract}

\keywords{
binaries: general --
infrared: stars --
open clusters and associations: individual ($\eta$ Cha) --
planetary systems: formation -- 
planetary systems: protoplanetary disks
stars: pre-main sequence
}

\section{Introduction}
At birth, protostars are surrounded by circumstellar 
(CS) gaseous disks that become the natural sites 
for planet formation. An important constraint for any planet formation scenario is 
the timescale available during which planet formation can occur, 
before the CS disks dissipate. 
Several studies have shown that near and mid-infrared excess flux, 
indicative of inner CS disks, rapidly declines with 50\% of 
low-mass stars losing their very inner dust disk within $\sim 3$~Myr
\citep[e.g.][]{Haisch01, Aurora06}.  The 
disappearance of inner disks seems to coincide with the end
of accretion, suggesting that both dust and gas disappear on 
similar timescales \citep{Gullbring98, Jayawardhana06}. 
The dissipation of the primordial disks in which gas planets could form seems to be completed
by $\sim 10$~Myr. However, CS disk evolution is currently poorly-constrained within the
$\sim$~5--10~Myr age range due to the lack of observationally well-characterized systems.

To study the influence of environment and stellar properties on
disk dissipation timescales, nearby young stellar cluster provide
ideal targets as many of their stellar properties, their ages and 
environment can be well constrained.  The nearby, 
coeval $\eta$~Chamaeleontis cluster at a distance of 97 pc and 
an isochronal age of $\simeq 8$~Myr \citep{Mamajek99, Lawson01,
Lyo04a, Luhman04} is an excellent laboratory 
to study disk evolution. The cluster consists of 
three early-type systems, and 15 low-mass stars with stellar types 
ranging from K6 to M5 \citep{Lyo04a, Luhman04}.  
\citet{Lyo03} showed that a substantial fraction of the $\eta$~Cha members have an
near-IR excess. This was confirmed by \citet{Megeath05} who observed
all late-type members at wavelengths between 3.6--8~$\mu$m with the Infrared 
Array Camera (IRAC) on the {\it Spitzer Space 
Telescope}, detecting disks around 40\% of the stars. Given the age of the $\eta$~Cha cluster,
this large disk fraction seems inconsistent with the $\sim$6~Myr maximum lifetime of inner disks
derived by \cite{Haisch01} from $L$-band observations of younger clusters. This argues against a 
single dissipation timescale in all environments and/or for all disk radii.

In this paper, we will employ the 
longer wavelength capabilities of the Infrared Spectrograph 
\citep[IRS;][]{Houck04} onboard {\it Spitzer\,} to firmly establish 
the census of disks for the 15 low-mass K- and M-type 
members of the cluster.  We then determine and discuss the influence of binarity on the CS
disk dispersal timescales, and on the specific angular momentum distribution
of the cluster members.

\section{{\it Spitzer\,} IRS observations of $\eta$ Cha}

We obtained $7.5-35$ $\mu$m low-resolution ($R = 60-120$)
spectra of the $\eta$ Cha cluster members with the 
IRS spectrograph. Integration 
times were set such that 
stellar photospheres could be detected with a signal-to-noise ratio $> 5$ across most of 
the IRS bandpass.  Our spectra are based on the
{\tt droopres} products processed through the S13.2.0 
version of the {\it Spitzer\,} data pipeline.  Partially based on 
the {\tt SMART} software package \citep{Higdon04}, our data was
further processed using spectral extraction tools 
developed for the "Formation and Evolution of Planetary Systems" 
(FEPS) {\it Spitzer\,} science legacy team.  
The spectra were extracted using a 6.0 pixel and 5.0 pixel 
fixed-width aperture in the spatial dimension for the observations 
with the first order of the short- ($7.5-14$ $\mu$m) 
and the long-wavelength ($14-35$ $\mu$m) 
modules, respectively. The background was subtracted 
using associated pairs of imaged spectra from the two nodded positions 
along the slit, also eliminating stray light contamination and anomalous dark 
currents. Pixels flagged by 
the data pipeline as being "bad" were replaced with 
a value interpolated from an 8 pixel perimeter surrounding the errant 
pixel. The low-level fringing
at wavelengths  $>20$~$\mu$m was removed using the 
{\tt irsfinge} package \citep{Fred03}.
The spectra are calibrated using a spectral response function derived 
from IRS spectra and Cohen stellar models for a suite of calibrators 
provided by the {\it Spitzer\,} Science Centre. 
To remove any effect of pointing offsets, we matched orders 
based on the point spread function of the IRS instrument, 
correcting for possible flux losses. The relative errors between 
spectral points within one order are dominated by the 
noise on each individual point and not by the calibration.  We 
estimate a relative flux calibration across an order of 
$\approx 2$~\% and an absolute calibration error between 
orders/modules of $\approx 5$~\%.

Our observations are sensitive to CS disks with substantial inner gaps 
not radiating at wavelengths $\leq 8$ $\mu$m as covered by IRAC. The maximum wavelength of 35 $\mu$m
of our spectra corresponds to a blackbody temperature of $\sim$90 K.  
Assuming a $\lambda^{-1}$ dependency for the dust opacity, and a
typical stellar temperature and radius of 3500 K and 1 $R_{\odot}$,
respectively, dust grains attain a temperature of 90 K at a radius 
of about 25 AU from the central star. Therefore our observations
should be sensitive to CS disks with a maximum inner gap radius of 
about this value.  Additionally, for circumbinary disks, 
\citet{Artymowicz94} argue that the inner edge of circumbinary 
disks will have a tidal truncation radius of $2.0-2.5\,a$.  Given 
a maximum detection radius of about 25 AU, we should be able to 
detect circumbinary disks in any $\eta$ Cha binary systems with
separations $2a\leq20$ AU.  

\begin{figure}
\centering
\includegraphics[width=0.5\textwidth]{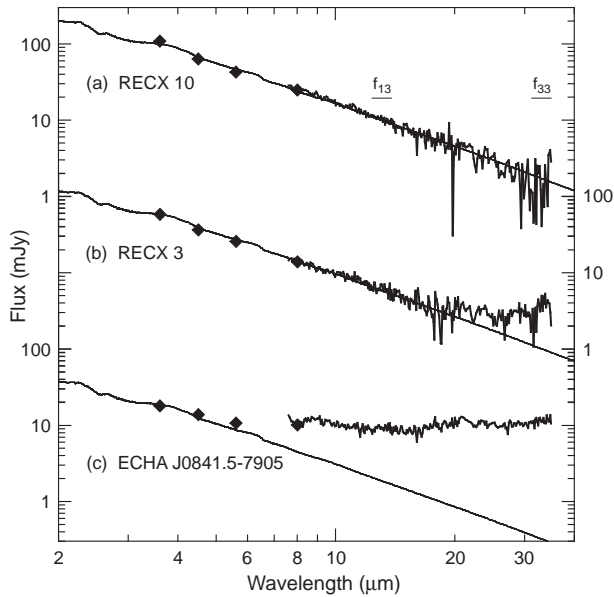}
\caption{Representative {\it Spitzer\,} IRS spectra of $\eta$ Cha
cluster stars.  Bold lines are the IRS spectra, diamonds are IRAC
photometry, while the thin lines are MARCS stellar models
generated using precise spectral types for the stars then scaled
to optical photometry. Panel (a) shows the diskless spectrum of
RECX~10.  Panel (b) shows the spectrum of RECX~3, a transitional
disk object with the weakest detected IRS disk with $f_{33}$ = 3.5 mJy.
Panel (c) shows ECHA J0841.5--7853 with a 10 mJy disk displaying
the typical spectral properties of a protoplanetary disk.  In panel (a)
we indicate the 13~$\mu$m and 33 $\mu$m filter bands used to
define the $f(33-13)$ colors in \S 2 and Table~\ref{Table1.tab}.
\label{Spectra.fig}}

\end{figure}

In Fig.~\ref{Spectra.fig} we show three representative IRS spectra. 
These examples demonstrate the excellent agreement between the 
IRS and IRAC data at 8 $\mu$m, both independently calibrated, and between the IRS spectra and the 
optical photometry-scaled stellar models for diskless systems.
From inspection of the spectra, we add RECX 3 and 4 to the 
disk census, meaning that 8 out of 15 (or about 50~\%) of 
low-mass $\eta$ Cha stars have IRS-detected disks. 
The newly detected disks lack excesses shortward of $\approx 15$~$\mu$m, 
consistent with the IRAC non-detection and are detected with 
$>5\sigma$ certainty at 33~$\mu$m.
To quantitatively test for the presence of disks, we calculated 
the $f(33-13)$ flux ratio derived from the IRS fluxes at 
13 $\mu$m and 33 $\mu$m, integrating over bandpasses of 1.3 $\mu$m 
and 3.3 $\mu$m centred at each wavelength, respectively.  
Table \ref{Table1.tab} lists the 13 
$\mu$m ($f_{13}$) and 33 $\mu$m ($f_{33}$) fluxes, and $f(33-13)$ 
colors for all the late-type $\eta$ Cha stars.  Our detection 
limits at 13 $\mu$m and 33 $\mu$m are $\approx 0.15$ and 0.3 mJy, 
respectively, showing the remarkable sensitivity of the IRS instrument. 
The six diskless late-type $\eta$ Cha stars with 33~$\mu$m 
detections of the stellar photosphere have a $f(33-13)$ color of $0.17 \pm 0.05$ (1$\sigma$).  
In contrast, the weakest disks detected in the transitional 
disk objects RECX 3 and 4 lack 13~$\mu$m excesses and have a 
$f(33-13)$ color of $\simeq 0.6$, while those stars with 
protoplanetary disks displaying significant disk excesses at 
both 13 and 33~$\mu$m have $f(33-13)$ colors exceeding $\simeq 1$.

\section{The influence of binarity on the disk fraction and the angular 
momentum
\label{F33.sec}}

Among the 15 late-type stars, six are binaries
with projected separations $\leq 20$~AU (column~2, Table~1).
Colour-magnitude diagram placement shows 
nearly half of the late-type stars are elevated by $0.5-0.7$ 
magnitudes compared to other cluster members of comparable spectral 
type, indicative of binary systems with near-equal luminosity 
components \citep{Lyo04b}. 
Of these stars, RECX 1 and 9 are resolved with a projected spatial
separation of $\simeq 20$ AU, following speckle and AO imaging
surveys for close companions \citep{Kohler02, Brandeker06}.
RECX 12 is an unresolved binary with an inferred separation of $\approx 4$ AU 
\citep{Brandeker06}, and has a dual-periodicity (1.3 d 
and 8.5 d) light curve measured during a photometric survey for 
starspot-modulated variability \citep{Lawson01}. 
From $v$\,sin$i$ measurements of RECX 
12 \citep{Covino97, Jayawardhana06}, we assume the 1.3-d mode to be
associated with the primary.  RECX 7 is a 2.6-d period, 
dual-lined spectroscopic binary of mass ratio 2.3:1 and separation 
$\approx 0.1$ AU \citep{Lyo03}. Given the spatial resolution of the \citet{Brandeker06} study,
the projected binary separations for the ECHA J0836.2--7908 and J0838.9--7916 systems are $<$4 AU. 
The cluster has a deficit of wide binaries at projected separations
$> 20-50$~AU \citep[e.g.][]{Brandeker06}.

\begin{figure}
\centering
\includegraphics[width=0.5\textwidth]{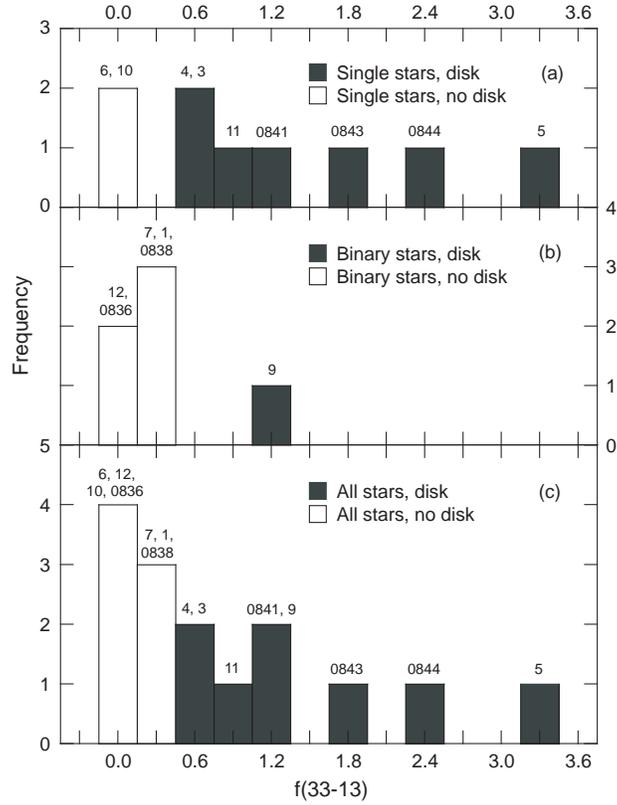}
\caption{Distributions of the $f(33-13)$
color for the late-type $\eta$ Cha stars. In this figure, individual 
stars identified by abbreviated labels, e.g. RECX 11 is labeled `11' 
and ECHA J0843.3--7905 is labeled `0843'.  Panel (a) and (b) plot 
the colors of single stars and binary stars with (without) disks, 
respectively, while in panel (c) we merge the $f(33-13)$ colors for 
the entire late-type population.
\label{F33.fig}}

\end{figure}

The disk frequency is summarized in Fig.~\ref{F33.fig}.
Our spectra reveal a remarkable divergence in disk presence 
as a function of binarity in the cluster's late-type population.
Of the eight detected disks, all but one are associated with single stars.
With nine single stars in total, about 80\%  have a CS disk.
Of the six known or suspected close binaries, only RECX 9AB  has a CS disk associated with it.
The presence of this disk, likely a circumprimary disk, was already inferred from ground-based $L$-band
photometry \citep{Lyo03}, H$\alpha$ spectroscopic measurements
of disk accretion \citep{Lawson04}, and IRAC photometry
\citep{Megeath05}. 
No circumbinary disks are detected. A non-parametric Mann-Whitney two-sample
$U$-test  indicates that the probability that the singles 
and binaries are drawn from the same parent sample is $P = 0.03$, 
i.e. CS disks are under-represented in $\eta$ Cha binaries with 
$> 2\sigma$ significance.  We surmise that the last "remaining" 
disk in a binary exists principally because the physical
separation of RECX~9AB greatly exceeds the projected $\simeq 20$ 
AU separation, given that no significant change in spatial separation or 
position angle was detected between the surveys of \citet{Kohler02} 
and \citet{Brandeker06}. The evolution of the disk around RECX~9A, therefore, 
effectively follows that of a disk in a single star system.
The $f(33-13)$ color of the RECX 9 disk 
suggests that the outer disk extends to at least 20 AU.  The
presence of an extended disk in combination with a circumprimary 
tidal truncation radius of $\approx 0.4\,a$ \citep{Artymowicz94} 
implies a physical binary separation $2a \geq 50$ AU.  If the evolution of RECX~9
is treated as a single star then a $U$-test gives $P = 0.003$, indicating that the evolution of single stars or binaries with separations equal or larger than that of RECX~9 is significantly different from the evolution of close binary systems.

To estimate a typical disk disappearance timescale, we assume a
rectangular probability distribution for the disk to dissipate 
around a mean timescale $t_{d}$ and width $2\sigma$. As constraints 
we assume that at $t = 0$ Myr all systems have
disks, and that by $t = 12$ Myr (the age of the $\beta$ Pic moving 
group) all disks have dissipated. 
We find for the binary and single star members, respectively, 
$t_{d} = 5 \pm 5$ Myr and $t_{d} = 9 \pm 3$ Myr.
If for our widest binary 
system, RECX 9AB, single star evolution is more applicable, 
the derived $t_{d}$ for disks in close binary systems could be even shorter.

\citet{HerbstMundt05} analysed the rotational evolution of samples of solar-like
PMS stars. Their key conclusion was that around half of PMS stars lose
significant surface angular momentum in the first $5-6$~Myr owing to rotational
coupling between star and disk.  During this phase, spin-up of the star is
prevented as it evolves towards lower luminosity, because angular momentum is
transferred to the CS disk. The other half of PMS stars evolve at almost
constant surface angular momentum, having lost their disks within the first few
million years.  These two groups of PMS stars are believed to evolve to form the
slowly- and rapidly-rotating groups of young main sequence stars, respectively.

\begin{figure}
\centering
\includegraphics[width=0.5\textwidth]{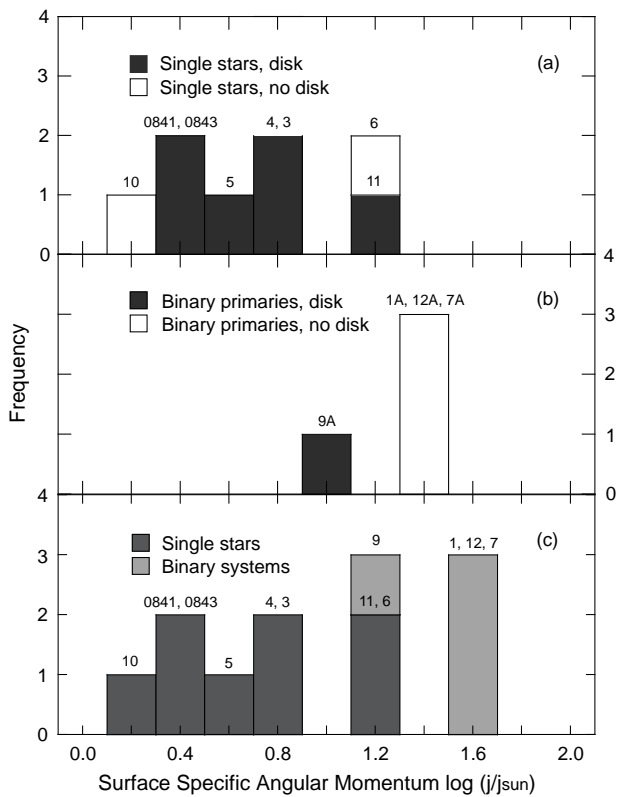}
\caption{Distributions of surface specific angular momentum $j$
in solar units. As in Fig. \ref{F33.fig}, individual stars are  
identified by abbreviated labels, with the addition that binary primaries are
further denoted, e.g. RECX 12A is labeled `12A'.  Panel (a) plots
$j$ values for single stars with (without) disks, while panel (b) is
a similar distribution for binary primaries. In panel (c) we assemble
the $j$ values for the cluster without compensation for binary
companions.  Panel (c) is the standard representation for the $j$
distribution where no knowledge is assumed about binary properties.
\label{SAM.fig}}
\end{figure}

We calculated the magnitude of the
specific surface angular momentum $j$ (surface angular momentum per 
unit mass) for the 12 late-type cluster members with rotation periods 
and photometry from \citet{Lawson01, Lawson02} and precise spectral 
types from \citet{Lyo04a}. We 
adopt dwarf temperature and bolometric correction scales to calculate
stellar luminosity and radii, with compensation applied for companion
stars within the binary systems based upon observed light and radial
velocity ratios. For binaries, we assume the observed rotation
period is associated with the primary star.
In Table \ref{Table1.tab} we express the derived $j$ values in solar 
units, where $j_{\odot} = 9.5 \times 10^{15}$ cm$^{2}$s$^{-1}$.  In 
Fig. \ref{SAM.fig} we plot the cluster's $j$ distribution for various
groupings of cluster objects. 
We see that {\it all\,} the high-$j$ systems are the primaries of 
$\eta$ Cha binary systems. With a mean $j = 22$ for binary primaries, 
this group differs significantly from the single stars with a mean 
$j = 6$.  A $U$-test gives a probability that the single stars and 
primaries are drawn from the same parent sample of only $P = 0.02$.  
The comparison would have been even more extreme if we had
not compensated for the presence of secondary stars.
For uncorrected binaries having a mean $j = 37$, the $U$-test gives a probability of $P = 0.004$.  

We thus provide novel empirical evidence for rotational disk locking, supporting models of star-disk magnetic coupling \citep[e.g.][]{Bouvier97, Kueker03}. 
We concur with \citet{HerbstMundt05} that high-$j$ stars are freed 
from any form of locking mechanism.  
In the $\eta$ Cha cluster, none of the high-$j$ stars have CS disks.
We additionally add the result that the high-$j$ tail in $\eta$ 
Cha consists entirely of the primaries of binary stars. 
This again argues that disk lifetimes in binary systems 
are shorter than those in single star systems. 
The $j$ samples of \citet{HerbstMundt05} are not corrected for binarity.  
From our results for the $\eta$ Cha cluster, we surmise that the 
high-$j$ tail of their distributions are populated 
with, and possibly dominated by, diskless binary 
systems.

\section{Conclusions}

The link between disk presence, angular momentum and 
binarity may have profound astrophysical importance in evaluating 
the time available for planet formation, as this is determined by the 
dissipation timescale of protoplanetary disks. 
The high angular momentum and paucity of disks in $\eta$ Cha binaries indicates a characteristic 
disk lifetime considerably less than the cluster's age of $\simeq 8$ 
Myr.  This is in contrast to the $\eta$ Cha singles, where the high 
disk fraction implies that their disk evolutionary timescale is 
comparable to, or slightly greater than, the age of the cluster.
Using a simple statistical approach, we estimate a mean disk dispersion timescale 
of $\sim 5$ Myr and $\approx 9$ Myr for close binary and single star systems, respectively.
Our results suggest that the correct evaluation of the {\it disk fraction\,} in PMS groups, 
and consequently the characteristic timescale for {\it disk longevity\,} 
during the PMS phase, critically depends on knowledge of binarity within 
a given PMS population. 

We cannot rule out that disk dissipation timescales could also depend
on stellar mass, as our sample only consists of late K- and M-type  
stars. Also, the derived timescales depend upon the detection wavelength,
reflecting that disk dissipation can have a radial dependence. However,
even if we limit our detections to those stars with IRAC excesses,
dissipation timescale for single stars only shortens by $\sim$10 \% to
$8 \pm 4$ Myr. In any case, our results imply that the assumption of a
single timescale for disk dissipation is not correct.  The rapid decline
of the disk fraction in the first few Myr as inferred by several  
near- and mid-IR studies, could be dominated by the dissipation of disks in close
binary systems. The slower dissipation of disks around single stars
could be the explanation for the long-lived disks seen in older PMS
clusters such as $\eta$ Cha.

The strong tidal torques exerted on disks in close binary systems will have a negative impact on the efficiency of planet formation \citep[e.g.][]{Kley01, Mayer05}.
Indeed, while $\sim$25 \% of exo-planets are detected in binaries, no planets have been
found in binaries with projected separations $< 20$ AU \citep{Raghavan06}, though at this point observational selection effects can not be ruled out.  Our results imply that in close binary systems the time available for planet formation is considerably shorter than in single star systems, which could severely inhibit the formation of planets. The apparent lack of planets in close binaries, therefore, might reflect the shorter disk dissipation timescale in binary systems. 

\acknowledgements  

EDF and AGGMT acknowledge support from {\it Spitzer\,} GO grant No. 3508 (PI WAL). 
WAL acknowledges support from UNSW@ADFA Faculty Research Grant
Programs 2005 and 2006. JB and ThH acknowledge support from the
EU HPR network contract No. HPRN-CT-2002000308.  We
thank Leen Decin, University of Leuven, for the 
stellar models.

\clearpage

\begin{deluxetable}{lcccccc}
\centering \tablecolumns{7} \tabletypesize{\small}
\tablewidth{0pt}

\tablecaption{Summary of binary, disk and rotational 
properties for late-type $\eta$ Cha stars \label{Table1.tab}}

\tablehead{
\colhead{Star} & \colhead{Binary} & \colhead{Disk} & 
\colhead{$f_{\rm 13}$} & \colhead{$f_{\rm 33}$} &
\colhead{$f(33-13)$} & \colhead{$j$} \\
\colhead{} & \colhead{} & \colhead{} & 
\colhead{(mJy)} & \colhead{(mJy)} & 
\colhead{} & \colhead{($j_\odot$)} \\
\colhead{(1)} & \colhead{(2)} & \colhead{(3)} & 
\colhead{(4)} & \colhead{(5)} & 
\colhead{(6)} & \colhead{(7)} 
}

\startdata
RECX 1             & 0.2\arcsec\  & n &  39.9 &   7.0 &  0.18 & 28.5 \\
RECX 3             &   & y &   6.1 &   3.5 &  0.57 &  5.0 \\
RECX 4             &   & y &  11.5 &   7.6 &  0.66 &  5.2 \\
RECX 5             &   & y &  21.0 &  68.0 &  3.24 &  4.2 \\
RECX 6             &   & n &   6.6 &   0.9 &  0.14 & 13.1 \\
RECX 7             & 0.001\arcsec\  & n &  27.2 &   6.9 &  0.25 & 21.0 \\
RECX 9             & 0.2\arcsec\ & y &  42.5 &  49.1 &  1.16 & 11.1 \\
RECX 10            &   & n &  10.4 &   1.1 &  0.11 &  1.9 \\
RECX 11            &   & y & 228.7 & 222.4 &  0.97 & 14.5 \\
RECX 12            & 0.04\arcsec\ & n &  16.0 &   2.0 &  0.13 & 25.2 \\
ECHA J0836.2--7908 & $<$0.04\arcsec\ & n &   1.7 &   0.0 &  0.00 & \nodata \\
ECHA J0838.9--7916 & $<$0.04\arcsec\ & n &   3.2 &   0.5 &  0.17 & \nodata \\
ECHA J0841.5--7853 &   & y &   9.0 &  11.3 &  1.25 &  2.7 \\
ECHA J0843.3--7905 &   & y & 155.8 & 267.8 &  1.72 &  2.5 \\
ECHA J0844.2--7833 &   & y &   7.2 &  16.6 &  2.30 & \nodata \\
\enddata

\tablecomments{Table columns:
Col 1.  Common star name.
Col 2.  Projected or inferred separations for the binaries (see \S 3.1 for details).
Col 3.  Presence of disk.
Col 4.  {\it Spitzer\,} IRS 13 $\mu$m flux $f_{13}$.
Col 5.  {\it Spitzer\,} IRS 33 $\mu$m flux $f_{33}$.
Col 6.  $f(33-13)$ color ratio = $f_{33}$/$f_{13}$.
Col 7.  Surface specific angular momentum $j$ in solar units,
where $j_{\odot} = 9.5 \times 10^{15}$ cm$^{2}$s$^{-1}$. For 
RECX 1, 7, 9 and 12, we list the $j$ value for the primary.
}

\end{deluxetable}

\clearpage

\clearpage

\clearpage

\begin{thebibliography}

\bibitem[Artymowicz \& Lubow (1994)]{Artymowicz94}
Artymowicz, P., \& Lubow, S. H.\ 1994, \apj, 421, 651

\bibitem[Bouvier et al.(1997)]{Bouvier97} Bouvier, J., Forestini, 
M., \& Allain, S.\ 1997, \aap, 326, 1023 

\bibitem[Brandeker et al.(2006)]{Brandeker06}
Brandeker, A., Jayawardhana, R., Khavari, P., Haisch, K. E.,
Jr, \& Mardones, D.\ 2006, astro-ph/0608352

\bibitem[Covino et al.(1997)]{Covino97}
Covino, E., Alcal\'a, J. M., Allain, S., Bouvier, J., 
Terranegra, L., \& Krautter, J.\ 1997, \aap, 328, 187

\bibitem[Gulbring et al.(1998)]{Gullbring98}
Gullbring, E., Hartmann, L., Briceno, C., \& Calvet, N.\
1998, \apj, 492, 323

    
\bibitem[Haisch et al.(2001)]{Haisch01}
Haisch, K. E., Jr, Lada, E. A., \& Lada, C. J.\ 2001, \apj, 
553, L153


\bibitem[Higdon et al.(2004)]{Higdon04}
Higdon, S. J. U. et al.\ 2004, \pasp, 116, 975

\bibitem[Herbst \& Mundt(2005)]{HerbstMundt05} 
Herbst, W., \& Mundt, R.\ 2005, \apj, 633, 967

\bibitem[Houck et al.(2004)]{Houck04}
Houck, J. R., Roellig, T. L., Van Cleve, J., Forrest, W. J.,
Herter, T. L., Lawrence, C. R., Matthews, K., Reitsma, H. J.,
et al.\ 2004, \apjs, 154, 18

\bibitem[Jayawardhana et al.(2006)]{Jayawardhana06}
Jayawardhana, R., Coffey, J., Scholz, A., Brandeker, A.,
\& van Kerkwijk, M. H.\ 2006, astro-ph/0605601

\bibitem[Kley(2001)]{Kley01} Kley, W.\ 2001, IAU Symposium, 
200, 511 

\bibitem[K\"ohler \& Petr-Gotzens(2002)]{Kohler02}
K\"ohler, R., \& Petr-Gotzens, M. G.\ 2002, \aj, 124, 2899

\bibitem[K{\"u}ker et al.(2003)]{Kueker03} K{\"u}ker, M., 
Henning, Th., \& R{\"u}diger, G.\ 2003, \apj, 589, 397

\bibitem[Lahuis \& Boogert(2003)]{Fred03}
Lahuis, F., \& Boogert, A. \ 2003 , in SFChem 2002: Chemistry as a Diagnostic of Star Formation, 
Edited by Charles L. Curry \& Michel Fich, NRC Press, Ottawa, Canada, 335.

\bibitem[Lawson et al.(2001)]{Lawson01} 
Lawson, W. A., Crause, L. A., Mamajek, E. E., \& Feigelson, 
E. D.\ 2001, \mnras, 321, 57

\bibitem[Lawson et al.(2002)]{Lawson02} 
Lawson, W. A., Crause, L. A., Mamajek, E. E., \& Feigelson, 
E. D.\ 2001, \mnras, 322, L29

\bibitem[Lawson et al.(2004)]{Lawson04}
Lawson, W. A., Lyo A.-R., \& Muzerolle J.\ 2004, \mnras, 
351, L39

\bibitem[Luhman \& Steeghs(2004)]{Luhman04}
Luhman, K. L., \& Steeghs, D.\ 2004, \apj, 609, 917

\bibitem[Lyo et al.(2003)]{Lyo03}
Lyo A.-R., Lawson, W. A., Mamajek, E. E., Feigelson, E. D., 
Sung, E.-C., \& Crause, L. A.\ 2003, \mnras, 338, 616

\bibitem[Lyo et al.(2004a)]{Lyo04a} 
Lyo A.-R., Lawson, W. A., \& Bessell M. S.\ 2004, \mnras, 
355, 363

\bibitem[Lyo et al.(2004b)]{Lyo04b} 
Lyo A.-R., Lawson, W.~A., Feigelson, E. D., \& Crause, 
L. A.\ 2004, \mnras, 347, 246

\bibitem[Lyo et al.(2006)]{Lyo06} 
Lyo A.-R., Song, I., Lawson, W. A., Bessell M. S., \&
Zuckerman, B.\ 2006, \mnras, 368, 1451

\bibitem[Mamajek et al.(1999)]{Mamajek99}
Mamajek, E. E., Lawson, W.A., \& Feigelson, E. D.\ 1999, 
\apj, 516, L77

\bibitem[Megeath et al.(2005)]{Megeath05}
Megeath, S. T., Hartmann, L., Luhman, K. L., \& Fazio, G. G.\
2005, \apjl, 634, 113

\bibitem[Mayer et al.(2005)]{Mayer05} Mayer, L., Wadsley, J., 
Quinn, T., \& Stadel, J.\ 2005, \mnras, 363, 641

\bibitem[Raghavan et al.(2006)]{Raghavan06} Raghavan, D., Henry, 
T.~J., Mason, B.~D., Subasavage, J.~P., Jao, W.-C., Beaulieau, T. D., \& Hambley, N. C.\ 2006, \apj, 646, 523 

\bibitem[Sicilia-Aguilar et al.(2006)]{Aurora06} 
Sicilia-Aguilar, A. et al.\ 2006, \apj, 638, 897 

\end{thebibliography}
\end{document}